\documentclass[aps,floats,showpacs,twocolumn,prl]{revtex4}
\usepackage{epsfig}
\usepackage{amssymb}
\usepackage[dvips]{color}
\usepackage[latin1]{inputenc}
\usepackage{graphicx}
\usepackage{epsfig}

\newcommand{\nc}{\newcommand}
\nc{\be}{\begin{equation}}
\nc{\ee}{\end{equation}}
\nc{\tcb}{}
\nc{\tcr}{}
\nc{\w}{{\rm w}} \nc{\s}{{\rm s}}

\begin{document}

\title{Describing Sr$_2$RuO$_4$ superconductivity in a
generalized Ginzburg--Landau theory}

\author{E. Di Grezia}
\email{digrezia@na.infn.it}
\affiliation{Universit\`a Statale di Bergamo, Facolt\`a di
Ingegneria, viale Marconi 5, I-24044 Dalmine (BG), Italy
and Istituto Nazionale di Fisica Nucleare, Sezione di Milano, via
Celoria 16, I-20133 Milan, Italy}
\author{S. Esposito}
\email{salvatore.esposito@na.infn.it}%
\affiliation{Dipartimento di Scienze Fisiche, Universit\`{a} di
Napoli ``Federico II'' and Istituto Nazionale di Fisica Nucleare,
Sezione di Napoli, Complesso Universitario di Monte S. Angelo, via
Cinthia, I-80126 Naples, Italy}
\author{G. Salesi}
\email{salesi@unibg.it} %
\affiliation{Universit\`a Statale di Bergamo, Facolt\`a di
Ingegneria, viale Marconi 5, I-24044 Dalmine (BG), Italy
and Istituto Nazionale di Fisica Nucleare, Sezione di Milano, via
Celoria 16, I-20133 Milan, Italy }

\

\begin{abstract}

\noindent We propose a simple explanation of unconventional
thermodynamical and magnetic properties observed for Sr$_2$RuO$_4$.
Actually, our two-phase model of superconductivity, based on
a straight generalization of the Ginzburg--Landau theory,
does predict two jumps in the heat capacity as well as
a double curve for the dependence of the critical temperature on an external
magnetic field. Such theoretical previsions well agree with the currently
available experimental data for Sr$_2$RuO$_4$.

\pacs{74.20.-z; 74.20.De; 74.70.Pq; 74.81.-g}

\end{abstract}

\maketitle

\noindent In a recent series of papers \cite{1,2,3,4} we have
succeeded in obtaining a straightforward generalizations of the
original Ginzburg--Landau (GL) theory \cite{GL} in order to
describe $s$-wave superconductors endowed with two critical
temperatures, or even spin-triplet one-phase superconductors. The
basic idea has been to introduce two different order parameters
represented by two charged scalar fields (really only one mean
field with two distinct gauge representations, see below)
describing Cooper pairs with electrons bound by a weaker or
stronger attractive force, respectively. The resulting theoretical
model is therefore able to describe superconductors with two
distinct superconducting phases, since the two order parameters
condensate, in general, at different critical temperatures.
Peculiar thermal and magnetic properties of these kinds of
superconductors have been discussed in \cite{2,3}. Here we only
mention that an additional discontinuity in the specific heat is
predicted, with respect to the conventional case, when passing
from a superconducting phase to the other one. Moreover, at low
temperature the London penetration length for the superconductors
considered is strongly reduced, and the Ginzburg--Landau parameter
$\kappa$ becomes a function of temperature. Instead, in
temperature region between the two phase transition, $\kappa$ is
constant and the system behaves as a type I or a type II
superconductors depending on the ratio between the two critical
temperatures. Such a ratio may be as large as $\sqrt{4/3}$
\cite{2,3} (that is, a maximum difference of $\sim 15\%$ between
the two critical temperatures) for very large self-interaction of
the Cooper pairs with respect to the electromagnetic coupling. By
allowing a suitable non-linear interaction among the two scalar
fields, the same theoretical model may as well account for
rotational degrees of freedom in superconductivity, that is
spin-triplet superconductors with a single phase (the two mutually
interacting order parameters condensate simultaneously at a same
temperature). In the corresponding model, the main thermodynamical
and magnetic properties of these $p$-wave superconductors turn out
to be essentially the same as for the conventional $s$-wave
superconductors.

All the above-seen properties have prompted us to explore the possibility to
use the proposed model in order to understand the intriguing
properties exhibited by Sr$_2$RuO$_4$ superconductors, which still
wait for a comprehensive and solid explanation (see, for example,
the review in \cite{6}).
The layered Sr$_2$RuO$_4$ is a superconductor with a very low
critical temperature and long coherence length. It possesses
pronounced unconventional features such as the invariance of spin
susceptibility (across its superconducting transition temperature
$T_c$), a strong dependence of $T_c$ on non-magnetic impurities,
evidence for two-component order parameter from field distribution
in the vortex lattice, and so on \cite{6}. All these features
strongly suggest that Sr$_2$RuO$_4$ superconductivity does not
involve standard $s$-wave singlet pairing but, rather, it could be
understood in terms of spin-triplet superconductivity. However,
the observation of a nodal structure of a superconducting gap with
a seemingly circular line node around a cylindrical Fermi surface,
in contrast to an expected nodeless $p$-wave superconductivity
in analogy to the case of superfluid $^3$He, together with other NMR
measurements, leads to a non definitive statement even about the
symmetry of the superconducting order parameter. In addition to these
features, strong (thermodynamic) evidence for a second superconducting
phase in Sr$_2$RuO$_4$ exists for high magnetic field applied along a
direction $H\parallel 100$. The formation of this second superconducting
state seems closely related to the upper critical field limit in a
parallel field configuration, as is seen in the temperature dependence
of $H_{c_2}(T)$ and in the scaling breaking of the field dependence
of thermal conductivity with $H_{c_2}$ \cite{6,Deg}.

Despite the failure in applying standard theoretical models to
describe such non conventional features of Sr$_2$RuO$_4$
superconductivity, the appearance of a second discontinuity in the
specific heat measurements seems to suggest the possible
application of our above-mentioned GL-like model with two charged
scalar fields \cite{2,3,4}. Note that, in a Ginzburg-Landau
framework, it is quite easy to introduce two or more ad-hoc order
parameters describing different phase transitions, just by
considering two or more scalar fields with different masses and
self-interaction coupling constants. In such a way, however, the
resulting model is not as predictive as the original GL theory,
since additional unknown physical parameters appear. In our
present model, instead, the introduction of additional degrees of
freedom, described by two complex scalar fields $\phi_\w$,
$\phi_\s$ (rather than a single one), \textit{does not involve new
unknown physical constants}, so that it is fully predictive. As a matter
of fact we do require that both scalar fields be endowed with
\textit{equal} bare masses $m$ and self-interaction coupling
constants $\lambda$. The Lagrangian describing the system is,
indeed, the following ($\hbar=c=1$)
\begin{eqnarray}
    {\cal L} &=& \left(D_{\mu }\phi_\w \right)^{\dagger }\left( D^{\mu }\phi_\w
    \right) + m^2\phi_\w^\dagger\phi_\w - \frac{\lambda}{4}(\phi_\w^\dagger\phi_\w)^2
    \nonumber\\
    &+&
    \left(D_{\mu }\phi_\s \right)^{\dagger }\left( D^{\mu }\phi_\s
    \right) + m^2\phi_\s^\dagger\phi_\s - \frac{\lambda}{4}(\phi_\s^\dagger\phi_\s)^2
    \nonumber\\
    &-& \frac{1}{4}F_{\mu\nu}F^{\mu\nu}\,.
    \label{Ellex2}
\end{eqnarray}
where \ $m^2>0$, \ $F_{\mu\nu}\equiv\partial_\mu A_\nu -
\partial_\nu A_\mu$ is the electromagnetic field strength, \ and \
$D_{\mu }\equiv\partial_{\mu }+2ieA_{\mu }$ \ is the covariant
derivative ($2e$ is the electric charge of a Cooper pair).
Different critical temperatures $T_1$ and $T_2$ are due to
different condensations of electrons in Cooper pairs, mediated by
different effective self-interaction (induced by loop
renormalization) \cite{2}. By starting with the normal state
system and lowering its temperature, we meet a first spontaneous
symmetry breaking (SSB) at a critical temperature $T_1$ and the
medium becomes superconducting (phase-I). By further lowering the
temperature at $T=T_2$, the condensation involving the other order
parameter is energetically favored and a new phase transition
starts (phase-II).
In a microscopic BCS-like framework, this corresponds to
assume that, as usual, at temperature $T_1$ electrons at the
Fermi surface (with energy $\epsilon_{\rm F}$) become bound into Cooper
pairs, and the BCS state $|\psi_{\rm BCS} \rangle$ describing this
condensate of electron pairs above the filled electron Fermi sea
is expressed as follows \cite{Ann}:
\begin{equation}
|\psi_{\rm BCS} \rangle \, = \, \prod_{\bf k} \left( u_{\w, {\bf
k}} \, c_{\w,- {\bf k} \downarrow} + v_{\w, {\bf k}} \,
c^\dagger_{\w,{\bf k} \uparrow} \right) |0 \rangle \, ,
\end{equation}
where $c^\dagger_{\w,{\bf k} \uparrow}$, $c_{\w,- {\bf k}
\downarrow}$ are the creation operators of electron and hole
states around (below) temperature $T_1$ with given momentum ${\bf
k}$, $-{\bf k}$ (single particle energy $e_{\bf k}$) and spin up,
down, respectively. In the standard BCS approximation, the
Hamiltonian of such a system is given by
$$
{\cal H} = \sum_{{\bf k}, \sigma} \left( e_k - \epsilon_{\rm F} \right)
c^\dagger_{\w,{\bf k} \sigma} c_{\w,{\bf k} \sigma} -
$$
\begin{equation}
 - |g^\w_{\rm
eff}|^2 \sum_{{\bf k}, {\bf k}^\prime} c^\dagger_{\w,{\bf k}
\uparrow} c^\dagger_{\w,-{\bf k} \downarrow} c_{\w,-{\bf k}^\prime
\downarrow} c_{\w,{\bf k}^\prime \uparrow}\,,
\end{equation}
where $g^\w_{\rm eff}$ is the effective electron-phonon
interaction coupling constant at temperature $T_1$. In the mean
field approximation, by assuming that each Cooper pair is much
larger than the typical spacing between particles, we can
introduce the expectation value
\begin{equation}
\Delta_\w = - |g^\w_{\rm eff}|^2 \sum_{{\bf k}^\prime} \langle
c_{\w,-{\bf k}^\prime \downarrow} c_{\w,{\bf k}^\prime \uparrow}
\rangle \, ,
\end{equation}
and the BCS Hamiltonian becomes approximately
$$
{\cal H} = \sum_{{\bf k}, \sigma} \left( e_k - \epsilon_{\rm F} \right)
c^\dagger_{\w,{\bf k} \sigma} c_{\w,{\bf k} \sigma} +
$$
\begin{equation}
 + \sum_{\bf k}
\left( c^\dagger_{\w,{\bf k} \uparrow} c^\dagger_{\w,-{\bf k}
\downarrow}  \Delta_\w + \Delta_\w^\ast c_{\w,-{\bf k} \downarrow}
c_{\w, {\bf k} \uparrow} \right) \,.
\end{equation}
This Hamiltonian may be easily diagonalized by a Bogoliubov
transformation with the introduction of creation and annihilation
operators $b^\dagger_{\w}$, $b_{\w}$ that are quantum
superpositions of electron and hole operators \cite{Ann},
obtaining
\begin{equation}
{\cal H} = \sum_{\bf k} \left( E_{\w , k} \, b^\dagger_{\w , {\bf
k} \uparrow} b_{\w , {\bf k} \uparrow} - E_{\w , k} \, b_{\w , -
{\bf k} \downarrow} b^\dagger_{\w , - {\bf k} \downarrow} \right)
\, ,
\end{equation}
the energy eigenvalues being given by $E_{\w \, k} = \sqrt{(e_k -
\epsilon_{\rm F})^2 + |\Delta_\w|^2}$. \\
\indent At a lower temperature $T_2$, some of the electrons at 
the Fermi surface already bounded in the formed Cooper pairs rearrange 
themselves to form new Cooper pairs with a different effective electron-phonon 
interaction $g^\s_{\rm eff}$ (see below), so that below $T_2$ the BCS state 
is given by
$$
|\psi_{\rm BCS} \rangle \, = \, \prod_{\bf k} \left( u_{\w, {\bf
k}} \, c_{\w,- {\bf k} \downarrow} + v_{\w, {\bf k}} \,
c^\dagger_{\w,{\bf k} \uparrow} + \right. 
$$
\begin{equation}
\left. +u_{\s, {\bf k}} \, c_{\s,- {\bf
k} \downarrow} + v_{\s, {\bf k}} \, c^\dagger_{\s,{\bf k}
\uparrow} \right) |0 \rangle \, ,
\end{equation}
with an obvious meaning for the quantities involved. This means
that an additional gap $\Delta_\s$ arises,
\begin{equation}
\Delta_\s = - |g^\s_{\rm eff}|^2 \sum_{{\bf k}^\prime} \langle
c_{\s,-{\bf k}^\prime \downarrow} c_{\s,{\bf k}^\prime \uparrow}
\rangle \, ,
\end{equation}
and the full Hamiltonian describing the system at temperature
below $T_2$ is then
$$
{\cal H} = \sum_{\bf k} \left( E_{\w , k} \, b^\dagger_{\w , {\bf
k} \uparrow} b_{\w , {\bf k} \uparrow} - E_{\w , k} \, b_{\w , -
{\bf k} \downarrow} b^\dagger_{\w , - {\bf k} \downarrow} + \right.
$$
\begin{equation}
\left. + E_{\s
, k} \, b^\dagger_{\s , {\bf k} \uparrow} b_{\s , {\bf k}
\uparrow} - E_{\s , k} \, b_{\s , - {\bf k} \downarrow}
b^\dagger_{\s , - {\bf k} \downarrow} \right) \, ,
\end{equation}
where $E_{\s \, k} = \sqrt{(e_k - \epsilon_{\rm F})^2 + |\Delta_\s|^2}$
are the energy eigenvalues for the novel kind of electron-hole
states. \\
\indent In the most transparent (for our purposes) GL formalism,
after the condensations due to the U(1) SSB, the mean total free
energy results as the sum of contributions from normal-conducting
electrons, and from weakly-coupled and strongly-coupled Cooper
pairs: \be
\begin{array}{l}
\displaystyle F = F_{\rm n}  \ \ \ \mbox{\rm for} \ \ \ T>T_1\,, \hfill \
\\ \ \\
\displaystyle F = F_{\rm n} + a_\w(T)\eta_0^2 +
\frac{\lambda}{4}\,\eta_0^4 \ \ \ \ \mbox{\rm for} \ \ 
T_2<T<T_1\,, \hfill \ 
\\ \ \\
\displaystyle F = F_{\rm n} + a_\w(T)\eta_0^2 +
\frac{\lambda}{4}\,\eta_0^4+\\ \hfill \ + a_\s(T){\chi_0}^2 +
\frac{\lambda}{4}\,\chi_0^4 \ \ \ \ \mbox{\rm for} \ \ \ T<T_2\,, \label{2}
\end{array} \label{I+II}
\ee
($\eta_0$ is the nonzero expectation value of $|\phi_\w|$; while
$\chi_0$ is the nonzero expectation value of ${\rm
Re}\{\phi_\s\}$), with
\be
a_\w(T) = - m^2 + \frac{\lambda +
4e^2}{16}\,T^2 \,, \quad a_\s(T)
= - m^2 + \frac{\lambda + 3e^2}{12}\,T^2 \,.
\label{3}
\ee %
Note that, despite the fact that the bare masses and
self-interaction coupling constants are the same for both scalar
fields, the GL effective parameters $a_\w(T)$, $a_\s(T)$ are not:
this depending on a different choice for the degrees of freedom, 
described through the scalar fields $\phi_\w$, $\phi_\s$, which 
undergone condensation \cite{1,2,param}. 
In the BCS formalism this corresponds to the fact
that, at lower temperatures, the electron-phonon effective
interaction may be different than at higher temperatures, due to
particular constitutive properties of the material considered.
In fact, even though the bare value of the electron-phonon coupling
constant $g_0$ does not change with temperature, the
effective coupling $g_{\rm eff}$ entering the BCS Hamiltonian
does exhibit such a behavior when introducing
temperature-dependent radiative corrections, the size of the
effect depending on the particular system considered. The net
result is the appearance of two critical temperatures, which are
defined by the vanishing of $ a_\w(T) $ and $ a_\s (T)$, that is
\be T^2=T_1^2=\frac{16m^2}{\lambda+4e^2} \,, \qquad\quad
T^2=T_2^2=\frac{12m^2}{\lambda+3e^2}\, \label{eq4}
\ee %
A peculiar prediction of this model is that two discontinuities in
the specific heat are expected at temperature $T_1$ and $T_2$
\cite{2}, the first one being due to the formation of Cooper 
pairs of the first kind, while the second one corresponding to the 
rearrangement of some electrons in Cooper pairs of the second kind. 
However, in contrast with other models with two independent
order parameters such critical temperatures may differ at most of
$\sim 15\%$ as clear from Eqs.\,(\ref{eq4}) \be
1<(T_1/T_2)^2<4/3\label{5}\,, \ee the value of this ratio
depending only on the ratio $\lambda/4e^2$ of coupling constants
of the scalar self-interaction and the electromagnetic
interaction. Note that, in general, the arrangement of the Cooper 
pairs at temperature $T_2$ does not automatically guarantee a visible 
second jump in the specific heat exactly at $T_2$, since this process can 
in principle be smoothed over some temperature interval. It is, 
however, remarkable that two jumps in the heat capacity are just 
observed for Sr$_2$RuO$_4$ in the presence of an external magnetic 
field. In fact with reference, for example, to the measurements 
reported in \cite{6,Deg}, for $H>1.2 T$ a second peak
in $C_e(T)$ appears at a second critical temperature near the
upper limit in (\ref{5}). The exact values of these transition
temperatures depend on the magnetic field applied (they decrease
with increasing magnetic field) and the effect seems to be
suppressed at very low temperature, but notably \textit{the ratio
of the critical temperatures appears always close to the upper
limit} $\sqrt{4/3}$.

If the present model will be further confirmed by future
experiments on other peculiar predictions (see \cite{2,3}), a
possible, general interpretation of Sr$_2$RuO$_4$ superconductivity
is the following. The striking features of this material strongly
depend on the dynamics of two different Cooper pairs formed in it,
probably due to peculiar geometric arrangement of atoms in the
compound considered \cite{6}. The different effective masses of
these pairs (see Eqs.\,(\ref{3})) are responsible of the two peaks
in the specific heat, corresponding to the condensation of
electrons in the two different Cooper pairs, which differ one
another from the relative effective bound of the pairs of
electrons. The value of the ratio of the two critical temperatures
close to the upper limit in (\ref{5}), then, points out the
extremely large value of the scalar self-interaction coupling
$\lambda$ with respect to the electromagnetic coupling (see Eqs.\,(\ref{eq4})).

In the present model it is assumed that such a scenario comes out
for electrons grouped in $s$-\textit{wave} pairs \cite{2,3}.
However, all the known phenomenology which has been ascribed to
$p$-\textit{wave} coupling \cite{6} is seemingly not in
contradiction with the last statement. In fact in Ref. \cite{4} we
have shown that by switching on a suitable non-linear interaction
potential among two different Cooper pairs ($\theta_1$, $\theta_2$
are the phases of the fields $\phi_1$, $\phi_2$)
\begin{equation} \label{v12}
V(\phi_1,\phi_2)\equiv
-\frac{\lambda\phi_0^4}{8}\ln^2{\frac{\phi_1^{\vphantom{\star}}}{\phi_1^\star}
\frac{\phi_2^\star}{\vphantom{\phi_2^\star}\phi_2^{\vphantom{\star}}}}
= \frac{\lambda\phi_0^2}{2}(\theta_1-\theta_2)^2
\end{equation}
the system behaves as a spin-triplet superconductor, with
thermodynamical and magnetic properties similar to the corresponding
ones of the $s$-wave model (\ref{Ellex2}).
In this way, our model is able to reconcile different,
unconventional features of Sr$_2$RuO$_4$ superconductivity.

\

\noindent Before concluding, we would address also another
interesting feature reported in the literature, that is the
``splitting'' of the $T_c(H)$ curve \footnote{Usually, in
Sr$_2$RuO$_4$ literature, the splitting is referred to the $H_c$
dependence on the critical temperature(s) $T_c$. However, while
for the present purpose the two complementary views are identical,
we prefer to think at the physical situation of the variation of
the critical temperature(s) induced by the applied magnetic
field.}. For sufficiently high magnetic fields and low
temperatures the first part of that curve is ordinarily
approximated for small fields to a straight line in a GL-like
framework \cite{Ann}. However, as results from the previous
discussion, for Sr$_2$RuO$_4$, it is crucial the condensation of
large fields \cite{6,Deg}, where the $T_c(H)$ curve can no longer
be approximated to a straight line. In such a case, higher
eigenvalues (labelled by an integer number $n>0$) corresponding to
Landau level solutions of the linearized Ginzburg--Landau equation
\cite{Ann} come into play in the equation determining the
depression in transition temperature as a function of the magnetic
field. Always referring to $s$-wave model (1), we parameterize
this equation as follows (for further details see \cite{Ann}): \be
T_{cn}(H) \simeq T_{0n} - \alpha_n(2n+1)\mu_0 H
\label{7}\label{Tcn} \ee where $T_{0n}$ an $\alpha_n$ are two
parameters both depending on the elementary flux quantum of vortex
lines and on the type (weaker or stronger bound electrons) of Cooper pair condensed,
while the integer $n$ labels the possible energy solutions of the linearized
GL equation. We have analyzed the available data on $T_c(H)$ for Sr$_2$RuO$_4$
(for definiteness, we took the data from \cite{6,Deg}) in the different regimes
(different $n$) pointed out by Eq.\,(\ref{7}), and the best-fit
estimate for the parameters are the following (see also Figure 1):
\begin{figure}
\begin{center}
\epsfig{height=5truecm,file=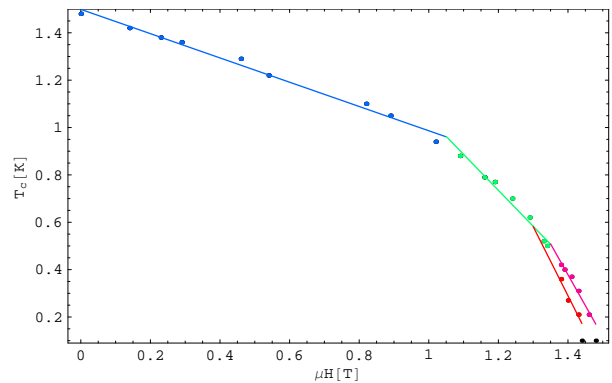}
\caption{Critical temperature versus magnetic field for Sr$_2$RuO$_4$
(best fit from the experimental data quoted in \cite{6,Deg})}
\end{center}
\label{fig1}
\end{figure}
\begin{eqnarray}
&&T_c^{n=0}=1.50 -0.51H\qquad\quad\!\textrm{for}\,\, H\leq 1.05 \nonumber \\
&&T_c^{n=1}=2.6 -1.52H\qquad\quad\,\,\textrm{for}\,\,1.05\leq H \leq 1.35 \nonumber \\
&&T_c^{n=2}= \left\{\begin{array}{ll}4.03 -2.61 H  &\quad\, \textrm{for
$H\geq 1.35$} \\ 4.38 -2.92 H & \quad\, \textrm{for $H\geq 1.30$}\label{n2}
\end{array}\right.
\label{8}
\end{eqnarray}

\noindent ($T_c$ and $H$ are given in K and T, respectively).
Notice that we can observe a sensible splitting only for $n=2$ eigenstates:
in (\ref{n2}) the first equation refers to weakly coupled 
Cooper pairs, the second equation to strongly coupled Cooper pairs.

Note that the ratios of the $H$ coefficients in (\ref{8}) are in good agreement
with the predictions of Eq.\,(\ref{7}): \ 2.98 versus 3, \ 5.12 and 5.72 versus 5
for $n=1\,/\,n=0$ and $n=2\,/\,n=0$, respectively. Though approximate (the approximation
of the true curve $T_c(H)$ with a set of straight lines is, of course, only indicative),
our analysis allows an immediate recognition of the trigger of the different regimes ruled
by different $n$ and gives some interesting information on the ``splitting'' of the
$T_c(H)$ curve. In fact from (\ref{8}) and, even more, from Fig.\,1 it is quite
evident that such a splitting happens during the transition from $n=1$ to $n=2$,
this transition starting at lower values of the applied magnetic field for weaker-coupled
Cooper pairs.
The $\alpha$ parameter in Eq.\,(\ref{Tcn}) is, indeed, inversely proportional to the
derivatives, with respect to temperature, of coefficients $a$ in (\ref{I+II})
evaluated at the respective critical temperatures
\cite{Ann}: %
\be %
\alpha = \frac{e \hbar}{m_\ast \dot{a}}
\ee %
(here, $m_\ast$ is the mass of the Cooper pairs), so that, from
Eqs.\,(\ref{3}) we easily get: %
\be %
\dot{a}_\w = \sqrt{e^2 + \frac{\lambda}{4}}\,, \qquad \qquad
\dot{a}_\s = \sqrt{e^2 + \frac{\lambda}{3}}\,.
\ee %
From these relations we deduce that $\alpha_\w \geq \alpha_\s$,
that is the slope of the $T_c(H)$ curve for the weak phase is
greater that the corresponding one for the strong phase. In
particular, from Eqs.\,(\ref{eq4}), the ratio of these two slopes
is given exactly by
the ratio of the two critical temperatures: %
\be %
\frac{\alpha_\w}{\alpha_\s} = \frac{\dot{a}_\s}{\dot{a}_\w} =
\frac{T_1}{T_2} .
\ee %
Thus, the measured slopes for $n=2$ give {\it independent} information on $T_1$ and $T_2$.
It is quite remarkable that the direct estimate of the ratio $T_1/T_2$ from
specific heat measurements \cite{6,Deg} is in good agreement with the just mentioned ratio
of the slopes of $T_c(H)$ for $n=2$ in (\ref{8}).

\

\noindent In conclusion, we have given some evidence that the unconventional
properties of Sr$_2$RuO$_4$ superconductivity may be accounted for
by a suitable generalization of the GL theory, as given in Refs.
\cite{1,2,3,4}. In such a model the two peaks appearing in the
specific heat curve at sufficiently high magnetic field applied
(and at low temperatures) correspond to the condensation of two
different types of Cooper pairs, where electrons are weaker or
stronger bound probably due to the peculiar geometry of
Sr$_2$RuO$_4$ crystals, composed by layers orthogonal to the magnetic field.
The ratio of the two critical temperatures, ruled by the ratio
of the scalar self-interaction coupling (ruled by $\lambda$) versus the
electromagnetic coupling (ruled by $e^2$), is in fact close to the upper
limit $\sqrt{4/3}$ of our model, meaning that the self-interaction is
in Sr$_2$RuO$_4$ much larger than the electromagnetic interaction. This may be directly
inferred from the specific heat measurements, but it is also in
very good agreement with an independent estimate from the slopes
(for $n=2$) of the ``splitted'' $T_c(H)$ curves in the scenario
envisaged here. The weaker bound Cooper pairs start the transition
from $n=1$ to $n=2$ at a lower value of the applied magnetic field
and at a higher temperature, as it is likely expected.

It is also quite interesting that a trough generalization of the proposed model,
accounting for a simple non-linear interaction potential among Cooper pairs
of different kinds, allows to describe a spin-triplet superconductor system, thus
recovering even all the unconventional features of Sr$_2$RuO$_4$,
usually interpreted in terms of $p$-wave pairing.

Although further experimental and theoretical work is desirable
for a comprehensive understanding of Sr$_2$RuO$_4$
superconductivity, the results illustrated here seem highly
promising to achieve such a goal.

\vspace{1cm}

\noindent {\large{\bf Acknowledgments}}

\vspace*{0.2cm}

\noindent Special thanks are due to C. Baldassari for stimulating discussions and
kind collaboration.


\begin{thebibliography}{}

\bibitem{1}
E. Di Grezia, S. Esposito and G. Salesi, Physica {\bf C451}, 86
(2007)

\bibitem{2}
E. Di Grezia, S. Esposito and G. Salesi, Physica {\bf C467}, 4
(2007)

\bibitem{3}
E. Di Grezia, S. Esposito and G. Salesi, Physica {\bf C468}, 883
(2008)

\bibitem{4}
E. Di Grezia, S. Esposito and G. Salesi, Mod. Phys. Lett {\bf
B22}, 1709 (2008)

\bibitem{GL} V.L.Ginzburg and L.D.Landau, Zh. Eksp. Teor. Fiz. {\bf 20}, 1064 (1950)

\bibitem{6} Mackenzie, A.P. and Y. Maeno, 2003, Rev. Mod. Phys 75, 657

\bibitem{Deg}  H. Yaguchi, K. Deguchi, M. A. Tanatar, Y. Maeno and T. Ishiguro,
J. Phys. Chem. Solids {\bf 63}, 1007 (2002); \ K. Deguchi, M. A.
Tanatar, Z. Mao, T. Ishiguro and Y. Maeno, J. Phys. Soc. Japan
{\bf 71}, 2839 (2002)

\bibitem{Ann}
J.F. Annett,  {\em Superconductivity, Superfluids and
Condensates}, (Oxford University Press; Oxford, 2003)

\bibitem{param}
E. Di Grezia, S. Esposito and G. Salesi, {\em Dependence of
the critical temperature on he Higgs field reparametrization},
arXiv:hep-ph/0812.0955 

\end{thebibliography}
\end{document}